\documentclass[conference]{IEEEtran}
\usepackage{fancyhdr}

\usepackage{cite}
\usepackage{amsmath,amssymb,amsfonts}
\usepackage{color,soul}
\usepackage{algorithmic}
\usepackage{graphicx}
\usepackage{textcomp}
\usepackage{xcolor}
\usepackage{float}
\usepackage{multicol}
\usepackage{enumitem}
\usepackage{titlesec}

\def\BibTeX{{\rm B\kern-.05em{\sc i\kern-.025em b}\kern-.08em
    T\kern-.1667em\lower.7ex\hbox{E}\kern-.125emX}}
\begin{document}

\title{From Architecture to Binary: Ensuring Cross-Domain Consistency in Model-Based Airborne Software Development}

\author{
Nils Schlautmann\textsuperscript{1},
Viktor Sinitsyn\textsuperscript{2},
Benjamin Engelhard\textsuperscript{3},
Florian Holzapfel\textsuperscript{4} \\
\textit{Institute of Flight System Dynamics, Technical University of Munich, Germany} \\
\texttt{nils.schlautmann@tum.de, viktor.sinitsyn@tum.de},\\ 
\texttt{benjamin.engelhard@tum.de, florian.holzapfel@tum.de}
}

\maketitle
\thispagestyle{fancy}

\begin{abstract}

This paper presents an airborne software development approach for manned and unmanned aerial vehicles aimed at reducing inconsistencies across system, model-based functional, and embedded software domains. In environments influenced by standards such as ARP-4754B and DO-178C, these inconsistencies typically stem from insufficient enforcement across domain boundaries rather than missing process definitions. Building on a previously proposed toolchain centered on a relational interface database, we identify recurring failure modes and propose a repository-centered implementation to address them, tailored to small, resource-constrained teams operating without heavyweight process overhead. Each domain is assigned a primary repository with cross-repository references and dedicated CI pipelines that generate, update, and validate the exchanged artifacts. Automated interface updates, differential change notifications, and consistency checks propagate changes with minimal manual effort and surface inconsistencies before the time-consuming code-generation and compilation steps. An initial implementation in an ongoing experimental project is described, with qualitative feedback from its early use.

\end{abstract}

\section{Introduction}\label{section:introduction}

Recent shifts in the aerospace market have led to a rapid increase in new entrants to the field of unmanned and experimental aerial vehicles. In this environment, short time-to-market, early demonstrations, and the ability to scale development within resource-constrained teams have become increasingly critical. At the same time, model-based functional software development has gained significant traction, often resulting in a separation between embedded software and functional development teams. While this separation enables specialization and scalability, it introduces substantial challenges to coordinate across these abstraction layers. These challenges are multiplied in growing development teams and in projects with rapidly evolving system architectures driven by new features or continuous feedback from ground or flight testing.

In~\cite{Sinitsyn.StreamlinedAirborneSoftware}, a streamlined, model-based software development process for large UAVs is presented for small to medium-sized development teams. At the core of the process, the centralized database \emph{dBricks} is used as a Single Source of Truth to represent the airborne system and corresponding interfaces. From this database, function definitions and transport-layer specifications are exported as inputs to the different software development teams. To ensure efficiency in start-up or experimental environments while at the same time enabling future steps towards certification activities, the process focuses on automatable activities but also considers incremental upgrades to align with standards such as ARP-4754B and DO-178C.

In practice, function parameters (application input/outputs) updated at the system level may unintentionally diverge from functional models or become inconsistent with existing transport-layer code. Conversely, parameters introduced during functional development may not be propagated back and therefore remain undocumented at higher levels. These issues are further compounded by limited cross-domain understanding of artifact dependencies and constraints. For example, an embedded software developer may be unaware that a change in execution rate affects functional behavior, while a functional developer may commit a model without realizing that subsequent code generation will fail due to incompatible data-type definitions.

Such inconsistencies are typically not caused by a lack of process definition, but by the absence of mechanisms that continuously verify and enforce the correct application of processes across artifacts owned by different roles and driven by differing priorities. They often remain undetected until late stages of integration or testing, where identifying and correcting root causes becomes time-consuming and costly.

To address these challenges, this contribution proposes to implement the software development process presented in~\cite{Sinitsyn.StreamlinedAirborneSoftware} using a repository-centered approach combined with Continuous Integration (CI) pipelines. In particular, the proposed approach focuses on introducing lightweight, consistency-enforcing mechanisms rather than heavyweight process gates, where the novelty lies not in CI but in providing small teams targeting ARP-4754B/DO-178C compliance with tools to surface inconsistencies early on. At the heart of the approach lies the centralized database \emph{dBricks}, against which downstream artifacts are generated and continuously validated. For each device, three repositories are defined and maintained by the system, functional, and embedded domains (Section\ \ref{section:domainsAndResponsibilities}). Clearly defined repository interfaces and dedicated pipelines ensure cross-domain consistency with minimal manual effort.

While not intended to replace formal certification activities, the implementation supports certification-oriented practices by improving traceability and enforcing consistency across development domains. The primary benefit lies in the early identification and resolution of inconsistencies during software development, thereby reducing integration risk and accommodating the differing priorities of the disciplines involved.

\subsection{Related Work}\label{section:relatedWork}

The foundations of the development processes used in this work were established by Hochstrasser in~\cite{Hochstrasser.Modularmodelbaseddevelopment}, which introduced \emph{mrails}, a process-oriented build tool focused on the functional model domain. Designed to support modular development in alignment with RTCA DO-331, \emph{mrails} enables multiple functional developers to define build jobs and coordinate parallel work while checking and correcting consistency across different components of functional models. It represents the first steps toward an automation-capable, integrity-checking development environment for model-based airborne software.

Building on this foundation, Dmitriev et al.~extended the process to cover a broader scope of DO-178C and DO-331 objectives, leveraging \emph{mrails} and CI to automate the execution of model static analysis, simulation-based testing, unit testing, and static analysis of source code~\cite{Dmitriev.ALeanand}. The process remained focused on the functional development domain, with only textual requirements as the interface to the system level, and static analysis of embedded source code as a one-way input from the embedded domain. A related extension to system-level verification was presented by Panchal et al.~in~\cite{Panchal.DO178Csoftwaredevelopment}.

The scope of the toolchain was subsequently broadened to the system architecture level by Dollinger et al., who introduced an aircraft-level perspective and provided software requirements as inputs to the process defined previously~\cite{Dollinger.BeLeanHow}. While this work significantly advanced the coverage of the development process, a formalized and enforceable interface between the system and lower-level development domains remained an open gap.

A step toward closing this gap was taken by Rhein et al.~\cite{Rhein.AHolisticApproach}, who introduced an interface database acting as a Single Source of Truth for artifact generation across multiple domains. As outlined above, this concept was subsequently concretized by Sinitsyn et al.~\cite{Sinitsyn.StreamlinedAirborneSoftware} and combined with the workflows of~\cite{Dmitriev.ALeanand}. Complementary work by Schmiechen~\cite{Schmiechen.InformationManagementfor} addresses the information-management side of the same toolchain — requirements management, data centralization, traceability, and automated quality checks within the ALM platform (Siemens Polarion\textsuperscript{\textregistered}).

Beyond this body of work, which originates from the same research group, several independent contributions have explored related challenges in the aerospace domain. Abdo, Broehan, and Thielecke proposed a model-based approach for the continuous validation of IMA platforms using a persistently running Jenkins server, demonstrating the viability of automated, continuous design validation. However, their focus lies primarily on architectural validation and simulation rather than on cross-domain consistency across multiple contributors~\cite{Abdo.AModelBasedApproach}. Similarly, Baron and Louis proposed automating repetitive certification tasks in a DO-178C context, though without support for model-based development, which differs from the flight control context addressed here~\cite{Baron.Towardsacontinuous,Baron.Frameworkandtooling}.

The broader relevance of consistency management across domains in model-based development is underscored by a study from Jongeling et al.~\cite{Jongeling.Consistencymanagementin}, which examined thirteen industrial settings. The central finding is a tight coupling between the adoption of shorter development cycles and increasingly pressing consistency management challenges. In almost all settings, manual consistency management activities are the primary bottleneck preventing more continuous development. This frequently leads to inconsistencies between artifacts such as system architectures, software models, and code being tolerated rather than resolved. One of the surveyed cases specifically concerns avionics, where consistency between system architecture models and implementation is required at all times, yet checks remain manual. A complementary perspective is provided by Torres et al.~\cite{Torres.Asystematicliterature}, whose literature review of cross-domain model consistency checking concludes that existing model management strategies remain insufficiently mature, are heavily tool-dependent, and do not adequately support the co-evolution of models across domains.

Jongeling et al.\ present a more directly related contribution in~\cite{Jongeling.LightweightConsistencyChecking}, which proposes a lightweight CI-based framework that detects structural inconsistencies between heterogeneous specification and design models — demonstrated on SysML and Simulink\textsuperscript{\textregistered} — and reports them without attempting automatic resolution. In contrast to the present work, their scope is confined to pairwise multi-view comparison at the design level and does not address the chain from functional models to generated code, nor the cross-domain alignment between system, functional, and embedded artifacts required for item development under aviation processes.

Cadavid et al.~\cite{Cadavid.Improvinghardwaresoftware} address Interface Control Document (ICD) management in large systems of systems using a documentation-as-code (DaC) philosophy, without relying on model-based development. Their approach uses a CI/CD pipeline to enforce consistency between per-interface ICDs and the artifacts generated from them, taking a decentralized structural path compared to the centralized database used here.

The reviewed works establish the development processes, the Single Source of Truth concept, and CI-based consistency checking in adjacent contexts, but an implementation enforcing the defined cross-domain process in practice has yet to be presented — a gap this paper contributes to closing.
\section{Current Development Process and Observed Failure Modes}\label{section:rolesAndFailures}

\begin{figure}[ht]
    \centering
    \includegraphics[width=0.48\textwidth]{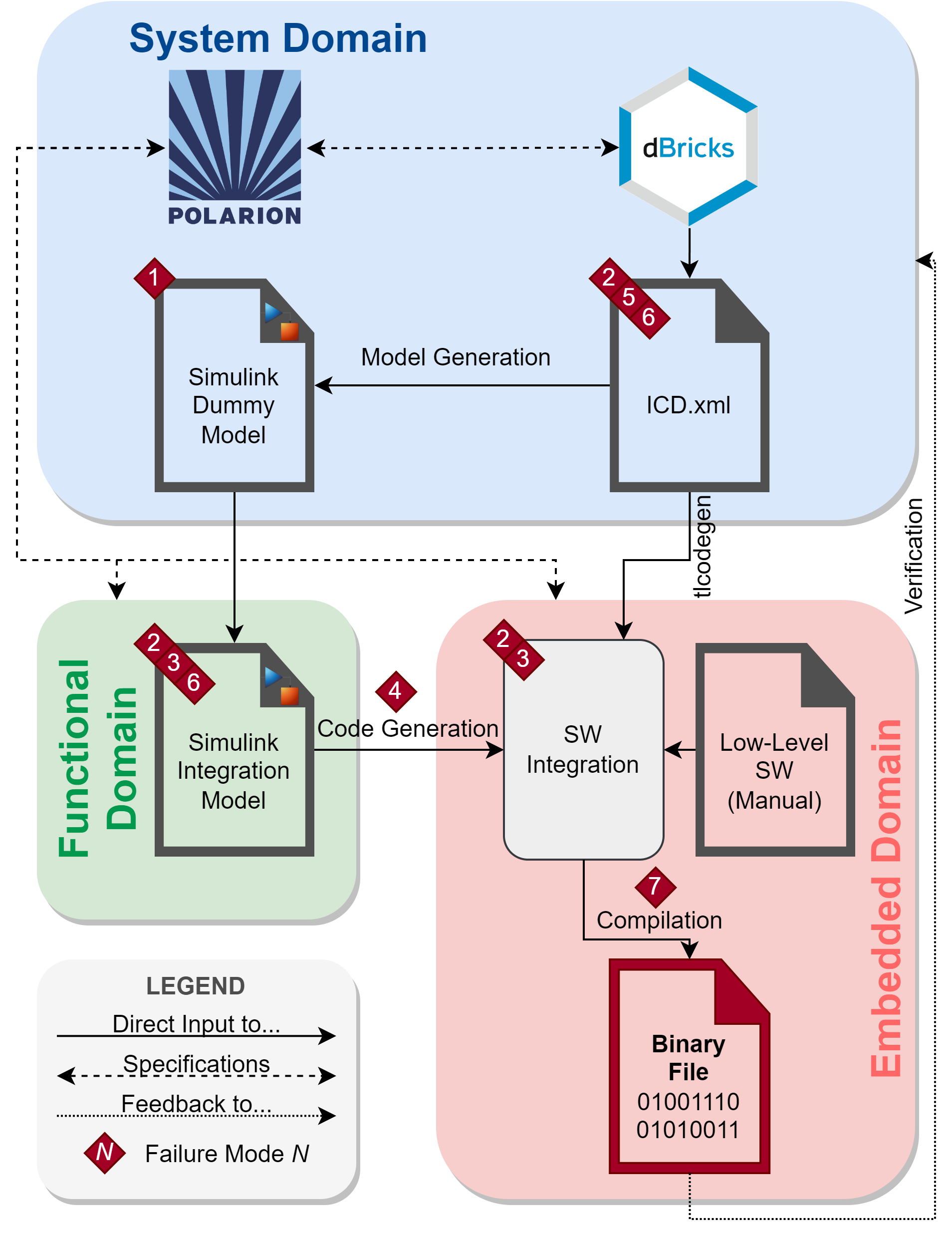}
    \caption{Development Domains, Relevant Artifacts and Observed Failure Modes}\label{fig:domains}
\end{figure}

\subsection{Domains and Responsibilities}\label{section:domainsAndResponsibilities}

When Sinitsyn's process~\cite{Sinitsyn.StreamlinedAirborneSoftware} is applied in practice at the Institute of Flight System Dynamics, three disciplines contribute to producing, maintaining, and exchanging the artifacts it defines. Within each discipline multiple engineers may contribute, and each has clear points of contact for the information exchange with the other domains. The three disciplines are described below and visualized in Figure~\ref{fig:domains}.

\textbf{System Domain:} The system domain is responsible for defining the architecture of the avionics system. Starting from the Concept of Operations (ConOps), system requirements are captured, aircraft-level functions are derived and allocated to items, and a physical architecture model is developed following the workflow described in~\cite{Dollinger.BeLeanHow}. After this initial development, the architecture is transferred manually into \emph{dBricks}, which subsequently serves as the Single Source of Truth for all downstream activities. Specifications and source data remain in Polarion\textsuperscript{\textregistered} and are linked to \emph{dBricks} for bidirectional traceability~\cite{Schmiechen.InformationManagementfor}. 

In typical projects at the institute, a large share of the components are Commercial-Off-the-Shelf (COTS) devices whose interfaces are fixed by the supplier; in these cases the team's contribution consists primarily of capturing the data consistently in \emph{dBricks}. However, for the few devices carrying custom software, the interfaces are designed in-house and require active coordination with the functional and embedded domains, especially for communication between different custom devices. The formal definition of the physical and logical interfaces by the systems team constitutes the central input for all downstream software development work. At the end of the development cycle, the system domain receives the compiled binaries and verifies them against system-level requirements during integration testing.

\textbf{Functional Domain}: The functional domain is responsible for developing the application-layer functionality and uses MathWorks Simulink\textsuperscript{\textregistered} as its primary authoring environment. Derived requirements are authored in Polarion\textsuperscript{\textregistered} based on ConOps and system requirements, and development follows the model-based process defined by Hochstrasser~\cite{Hochstrasser.Modularmodelbaseddevelopment}. Multiple team members contribute modularly to the functionality with the support of the in-house build tool \emph{mrails}. Simulation models and test cases are developed from system-level inputs, but are currently disconnected from the architecture model, which limits traceability between architectural and behavioral artifacts.

The flight software model is assembled by a designated integrator in an integration repository, which forms the boundary on both sides of the domain: it ingests the function parameters defined at system level and releases the integrated model to the embedded domain for code generation. A single device may host several integrated functions running at different rates, for example navigation and flight control on the same processor. Only the top-level function interface is drawn from \emph{dBricks} while the internal decomposition into submodels is managed within the functional domain itself.

\textbf{Embedded Domain:} The embedded domain develops the low-level software that runs on the target device, which at the institute is typically bare-metal or based on a Real-Time Operating System (RTOS). The domain's responsibilities cover drivers, the board support package (BSP), middleware, and the initialization of the different communication interfaces. The transport-layer code that encodes and decodes function parameters into protocol-specific messages is generated automatically using the tool \emph{tlcodegen} developed by Schwaiger~\cite{Schwaiger.EfficientSoftwareTools}. The transport-layer code templates consumed by this tool are device-specific and authored by the embedded team. For the generated code to remain consistent with the rest of the system, \emph{tlcodegen} must be invoked with inputs matching the released functional interfaces.

Once the low-level software is implemented and tested, the embedded team takes the released top-level functional model to generate the corresponding source code, integrates it with the low-level software and generated transport-layer code before compiling the binary that is then handed over to the system domain. Findings from system-level integration testing with the binary are fed back into the software and functional development process iteratively, closing the loop between the three domains.

\subsection{Artifacts Affected by Inconsistencies}\label{section:affectedArtifacts}

Previous works have already contributed mechanisms for ensuring consistency between a large number of artifacts in the development process described above,
including across domain boundaries. For example, the system and functional domains share the same ALM platform, which enables direct traceability between ConOps, system
requirements, and derived software requirements~\cite{Schmiechen.InformationManagementfor}. Other examples are the internal consistency checks inherent to the \emph{dBricks} database itself~\cite{Sinitsyn.StreamlinedAirborneSoftware} or the consistency ensured between functional modules by using \emph{mrails}~\cite{Hochstrasser.Modularmodelbaseddevelopment}.

In practice, however, it is less the \emph{specifications} than the \emph{design outputs} of the three domains that suffer from inconsistency in the fast-paced environment of the institute. The artifacts impacted by these issues were described in detail by Sinitsyn~\cite{Sinitsyn.StreamlinedAirborneSoftware} and are introduced here to outline where consistency between them may be lost.

\textbf{ICD XML (System Domain):} The ICD is exported from \emph{dBricks} as an XML file and serves as the single machine-readable description of the interfaces for a given device. It defines the existing physical ports, the transport layer, the integrated functions assigned, as well as the logical inputs and outputs of these functions~\cite{Sinitsyn.StreamlinedAirborneSoftware}. Different revisions of the ICD XML may remain compatible with one another at specific abstraction layers: a change that only affects the transport layer or reroutes physical ports, for example, leaves the function parameters untouched and does not require updates to downstream Simulink\textsuperscript{\textregistered} models. This layered compatibility is important for development speed, as it allows each domain to avoid unnecessary rework when a change does not propagate into its layer. As a consequence, an ICD revision is relevant to a different subset of the domains depending on its content: the system team assesses whether a revision can be shared downstream, the functional team only needs to react to changes affecting function parameters, and the embedded team is concerned with changes to the transport layer or to the function-to-device assignments.

\textbf{Simulink\textsuperscript{\textregistered} Dummy Model (System Domain):}
From the exported ICD XML, a Simulink\textsuperscript{\textregistered} dummy model is generated automatically and subsequently owned by the system team, consisting of a model file (\texttt{.slx}) together with the associated Simulink\textsuperscript{\textregistered} Data Dictionaries (\texttt{.sldd}). It contains only the inputs and outputs of the device's integrated functions and serves as the structural skeleton from which the functional models are developed. Although its generation is fully machine-driven, the dummy model itself carries no explicit reference back to the ICD XML revision from which it was derived.

\textbf{Simulink\textsuperscript{\textregistered} Integration Model (Functional Domain):} The Simulink\textsuperscript{\textregistered} integration model is obtained by adding the functional submodels to the dummy model. It is version-controlled in GitLab repositories and sub-repositories, which provides internal traceability. As with the dummy model, however, it carries no explicit link back to the ICD XML revision whose interface it was constructed against, so there is no in-tool mechanism to detect when the functional model and the current interface definition have become inconsistent.

\textbf{Embedded Software, Transport-Layer Code and Functional Source Code (Embedded Domain):} The embedded software is typically derived from prior projects, and board support packages can be pulled in from separate repositories when the same hardware platform is reused. Traditionally, the templates consumed by \emph{tlcodegen} to generate the transport-layer code have been developed on a per-project basis. The transport-layer code itself is generated directly from the ICD XML and stored together with the rest of the embedded software. In earlier project setups, the XML was additionally checked in alongside the embedded sources and updated to the latest revision as needed. Because the XML embedded alongside the low-level software is updated independently of the functional domain's release cycle, inconsistencies between the two can remain hidden until compilation. For the embedded software to scale across projects sharing the same hardware platform, both the templates and the low-level software components must be designed for reuse, which has so far been difficult to achieve consistently.

Across all design-output artifacts, the common gap is the absence of an explicit reference from each artifact back to the ICD XML revision it was built against. This missing link is what makes consistency between the design outputs brittle, and it is the problem addressed in this paper.

\subsection{Failure Modes Observed in Practice}\label{section:failureModes}

During the application of the toolchain described in~\cite{Sinitsyn.StreamlinedAirborneSoftware} to several UAV and manned-aircraft projects, a number of recurring failure modes emerged that stem from inconsistencies or inefficiencies in the mechanisms ensuring consistency between the artifacts introduced above. The observations reported below are drawn from concrete project experience and complement the process description in~\cite{Sinitsyn.StreamlinedAirborneSoftware}: they illustrate that, despite the availability of a Single Source of Truth in \emph{dBricks} and the use of machine-readable intermediate artifacts, cross-artifact consistency is neither automatic nor self-evident once multiple engineering domains operating at different iteration cadences interact. The failure modes are also marked in Figure~\ref{fig:domains}.

\begin{enumerate}[label=\textbf{FM\arabic*},leftmargin=*,align=left]
    \item \textbf{Absence of differential information for ICD updates:}
    Whenever the ICD XML is re-exported and the associated Simulink\textsuperscript{\textregistered} dummy model and data dictionary (\texttt{.sldd}) are regenerated, no structured change history is propagated to the functional domain. The only artifacts delivered to the functional team are the updated model and \texttt{.sldd} files themselves. Identifying what has actually changed — whether a function parameter has been renamed, a data type modified, or a signal added or removed — therefore relies on manual comparison, which is both error-prone and disproportionately time-consuming, even for minor updates.

    \item \textbf{Version drift between co-evolving XML exports:}
    Because the ICD XML can be exported from the Single Source of Truth at arbitrary points in time, the embedded and functional domains frequently operate against different XML revisions. In many cases this desynchronization is deliberate: changes confined to the transport layer can be adopted rapidly on the embedded side, whereas updates on the functional side typically require more manual effort and thus a longer iteration cycle. Decoupling the two processes is therefore beneficial for efficiency during development. Unfortunately, no mechanism is currently in place to determine whether two XML revisions are mutually compatible — for example, whether function parameter names, data types, or scaling factors have changed in a manner requiring updates to the Simulink\textsuperscript{\textregistered} models. As a consequence, incompatibilities only surface when transport-layer code is linked against generated functional source code. In practice, a failed build is preceded by a 20--60\,min Simulink\textsuperscript{\textregistered} build phase, with a corresponding loss of productive time.

    \item \textbf{Inconsistent units and scaling between transport layer and functional model:}
    In past projects, unit conventions, scaling factors, and data-type conversions were not systematically aligned between the transport layer and the Simulink\textsuperscript{\textregistered} functional model. For example, some actuators expect position commands at a resolution of $0.1^\circ$, while others accept only integer degrees. Historically, reconciling these differences was left to the functional domain and handled on a case-by-case basis. The resulting errors typically remained latent until system-level testing, where command outputs were found to deviate by orders of magnitude. Such late detection is particularly costly, as it entails full re-integration and re-testing cycles, and because the root cause may be easily confused with functional defects in control laws themselves.

    \item \textbf{Data-type inconsistencies within Simulink\textsuperscript{\textregistered}:}
    Erroneous manual updates carried over from the Simulink\textsuperscript{\textregistered} dummy model into the integration model (e.g.\ typo corrections) can introduce subtle data-type inconsistencies that remain latent in the functional model. These inconsistencies typically only surface later, when the embedded team attempts code generation or compilation, at which point the functional team may no longer be immediately available to diagnose the root cause.

    \item \textbf{Loss of ad-hoc ICD fixes due to regeneration:}
    During equipment bringup, minor corrections to database errors are occasionally applied directly to the ICD XML file that is stored alongside the embedded software. Because the XML resides in the same repository as the transport-layer code, such edits are quick to perform and tempting to commit locally rather than to propagate back to the system domain and into \emph{dBricks}. One factor contributing to this issue is that the generation of the XML file needs to be manually triggered by a member in the system domain. Consequently, the local XML and the \emph{dBricks} definition drift apart, and the next regular XML export silently overwrites the local fixes. Previously resolved issues then reappear, and substantial effort is spent locating the regression and identifying the original commit that had addressed it.

    \item \textbf{Undocumented parameters introduced at the functional level:}
    During model development, the functional team may introduce additional parameters in the Simulink\textsuperscript{\textregistered} model that have no counterpart in the ICD. Such parameters may imply the need for the system domain to derive, for instance, the definition of custom device-to-device communication in order to exchange the added parameters. With no automated means to highlight such additions or to feed them back to the system level, the resulting gap typically surfaces only when new interface requirements are consolidated manually, at which point both the ICD database and the affected models must be reworked.

    \item \textbf{Toolchain and environment inconsistencies across workstations:}
    Finally, significant time is regularly lost during integration and flight testing when a modified functional model has to be rebuilt, but the embedded team is not immediately available to do so. In principle, rebuilding is a local activity that any engineer with access to the model should be able to perform. In practice, differences in installed toolchain versions and configurations across engineers' workstations produce compilation errors that are difficult to diagnose under field conditions. This creates an organizational bottleneck around a small group of engineers with known-good environments and undermines the responsiveness that the otherwise highly automated toolchain is intended to provide.
\end{enumerate}

Overall, these observations indicate that the failure modes are not rooted in the automated generation steps themselves. Instead, issues frequently arise where there is an absence of lightweight mechanisms to surface, compare, and communicate change information across domain boundaries. Particularly, wherever transitions between automated and manual activities can be observed or artifacts evolve in parallel at different design cadences, such issues are compounded. To address the observed failure modes, a repository-centered approach with CI pipelines as the lightweight mechanisms to ensure consistency is proposed in Section~\ref{section:repoApproach}.
\section{Proposed Repository-Centered Approach}\label{section:repoApproach}

\begin{figure}[ht]
    \centering
    \includegraphics[width=0.48\textwidth]{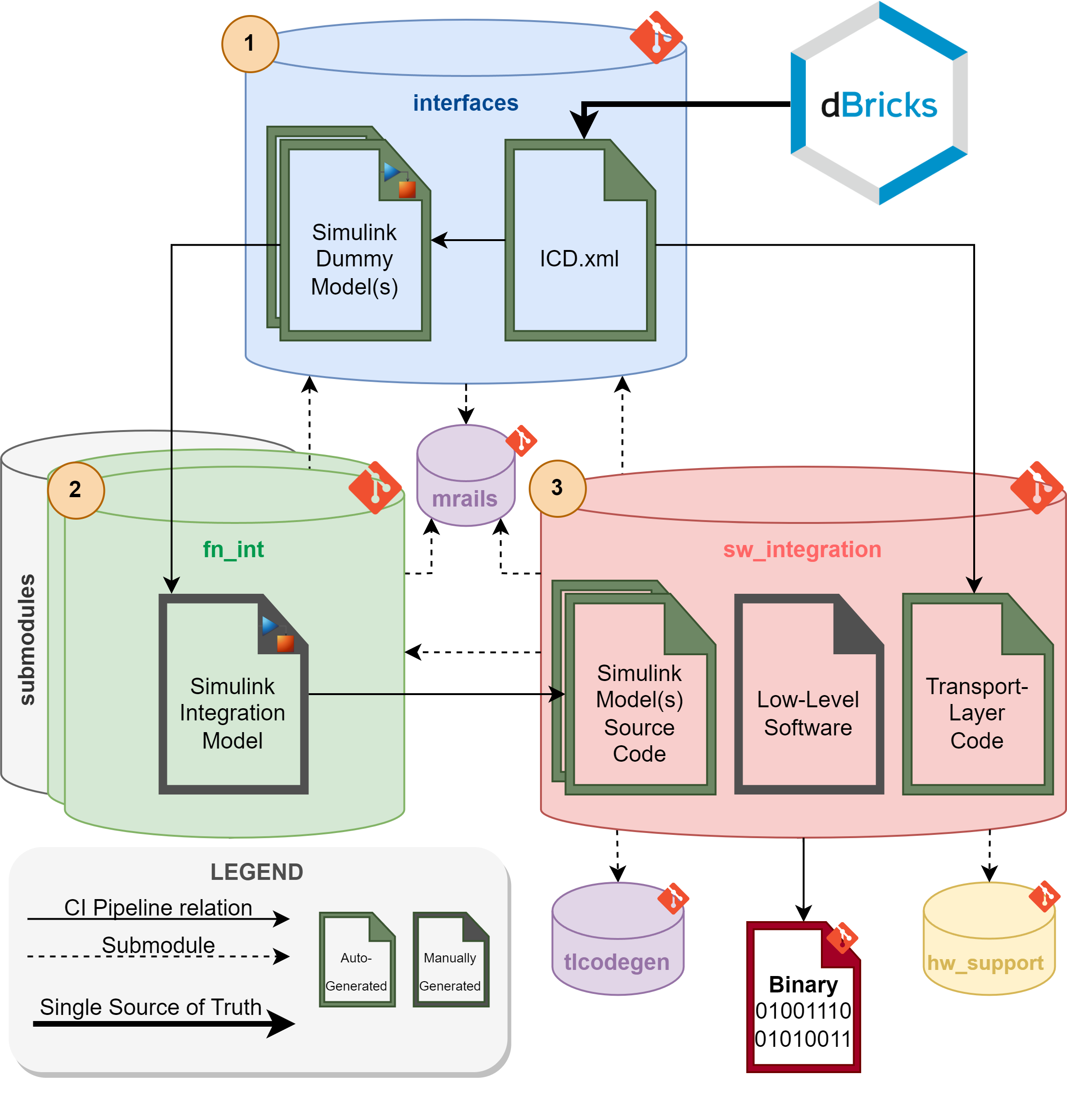}
    \caption{Proposed Repository Structure}\label{fig:repositories}
\end{figure}

To ensure cross-domain consistency for the artifacts identified in Section~\ref{section:affectedArtifacts}, each domain is assigned a primary repository that serves as its interface to the other domains. Each repository is complemented by a set of CI pipelines that enforce the processes governing the transitions between the three repositories by continuously detecting inconsistencies. This allows each domain to work independently while keeping all artifacts traceable and consistent. The proposed implementation is based on GitLab and its CI/CD tooling, extended with custom Docker-based runners.

To keep the process as efficient as possible, the repositories are granulated as coarsely as possible, while still ensuring that a change in one repository does not force changes in repositories it does not affect. The system and embedded domains each produce a device-level atomic artifact — the ICD XML and the compiled binary, respectively — and therefore maintain one repository per device. In the functional domain, two or more integrated functions may be allocated to the same device and hence compiled into the same binary. For example, navigation and flight control functions may run on a shared processor. To allow independent teams to develop these in parallel, each integrated function has its own repository. Because such functions may interface with one another and require matching interface definitions, the Simulink\textsuperscript{\textregistered} dummy models for all functions on a device are managed jointly within the system domain's interface repository.

A special role is taken by the interface repository, as it serves as the sole entry point from \emph{dBricks} into the development toolchain and exposes the artifacts relevant to downstream processes. This ensures that the system domain retains full control over the versions and content of the interface definitions that propagate across domain boundaries. Cross-repository references are implemented as Git submodules, so that each consuming repository is anchored to a specific revision of its upstream dependency, meaning that changes only take effect downstream once the reference is explicitly updated. The resulting structure is shown in Figure~\ref{fig:repositories}, and each repository is described in the remainder of this section.

Together, the proposed repository structure, pipelines, and submodule references address the failure modes identified in Section~\ref{section:failureModes}, as detailed in the following subsections.

\subsection{Interface Repository (System Domain)}\label{section:interfaceRepo}

\begin{figure*}[ht]
    \centering
    \includegraphics[width=0.99\textwidth]{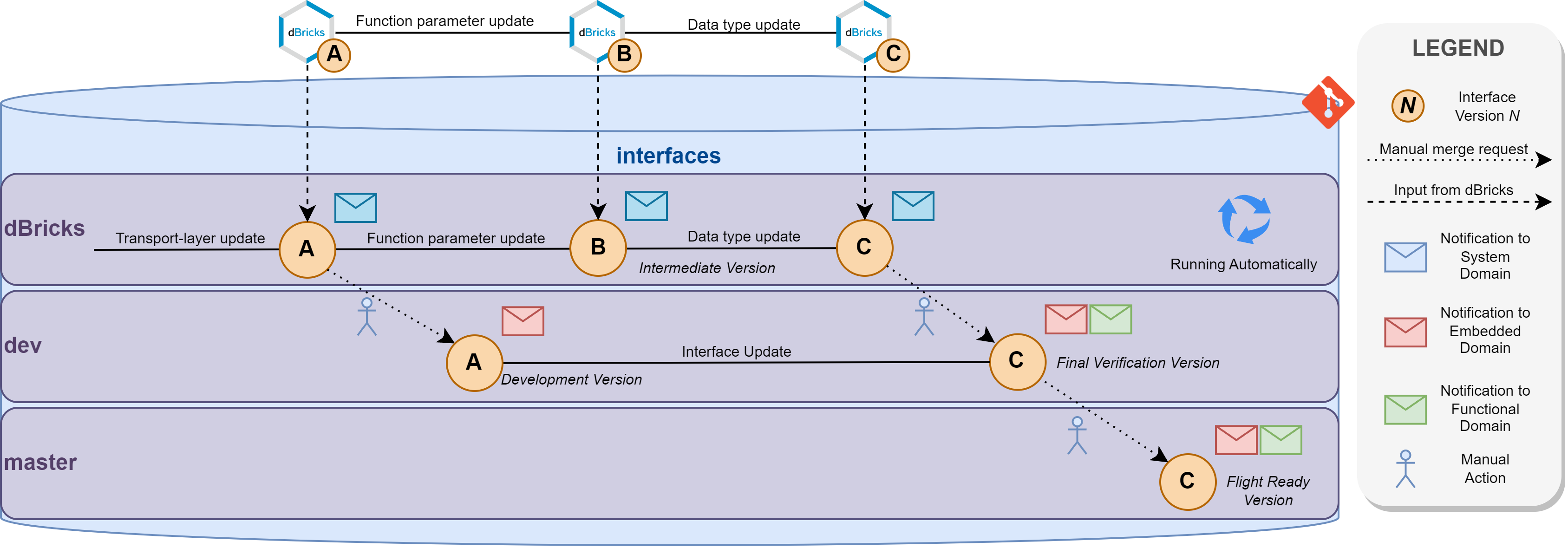}
    \caption{Interface Repository Branches and Pipelines}\label{fig:interfacePipeline}
\end{figure*}

The interface repository is the only bridge between \emph{dBricks} and the downstream development domains. It contains exclusively the artifacts required by those domains: the ICD XML, retrieved from \emph{dBricks} through its API, and the \emph{mrails}-compliant Simulink\textsuperscript{\textregistered} dummy models, generated from the XML by custom MATLAB scripts. Both artifacts are produced automatically by the repository's pipelines, so that the system domain's contribution is limited to controlling \emph{which} \emph{dBricks} revisions are released downstream and \emph{when}.

To make this release control explicit while still accommodating the different iteration cadences of the downstream domains, the repository is organized into three branches:

\begin{enumerate}
    \item \textbf{\texttt{dBricks}}: a mirror branch that pulls the latest state of the ICD from \emph{dBricks} and regenerates the corresponding Simulink\textsuperscript{\textregistered} dummy models. It allows the system domain to review proposed changes before any downstream team is exposed to them.
    \item \textbf{\texttt{dev}}: the development branch consumed by the functional and embedded teams. Updates from the \texttt{dBricks} branch are merged here through a manually triggered merge request once the system domain considers a revision ready for integration. Different commits on the \texttt{dev} branch may be consumed by the functional and embedded domains concurrently while still being mutually compatible. This compatibility is verified in the software integration repository (Section~\ref{section:swIntegrationRepo}) and provides the flexibility needed during early integration and testing.
    \item \textbf{\texttt{master}}: the branch used for flight testing. Functional and embedded releases must reference consistent commits on this branch. Merging \texttt{dev} into \texttt{master} is a formal system-domain decision and is again executed through a manually triggered merge request. This action constitutes the system domain's ``ready for flight'' statement for the interface definition.
\end{enumerate}

The central role of the pipelines is to keep the affected individuals informed about changes they need to respond to. On each commit, the new revision is compared against its predecessor and the detected changes are classified. In contrast to a purely syntactic diff, the comparison exploits the structured nature of the interface definition and assigns each change to one of five categories, each associated with the domain(s) that need to be notified:

\begin{itemize}
    \item \emph{physical ports} --- Embedded Domain,
    \item \emph{functions} --- Functional and Embedded Domain,
    \item \emph{function parameters} --- Functional and Embedded Domain,
    \item \emph{data types} --- Functional and Embedded Domain,
    \item \emph{transport layer} --- Embedded Domain.
\end{itemize}

Of these, the last three categories only affect artifacts that are automatically regenerated on the embedded side and therefore do not require manual action by the embedded domain beyond updating submodule references and re-running the corresponding pipeline. This distinction is carried into the notification stage, so that each stakeholder receives only the changes relevant to their role and low-impact modifications are filtered out.

Notifications are delivered by email, which minimizes dependencies on platform-specific tooling and supports communication with both internal and external stakeholders. GitLab's Service Desk functionality is not suitable for proactive triggering from within a CI pipeline, and a dedicated notification stage was therefore implemented in Python. The recipient groups are configured through the pipeline's configuration file. Changes on the \texttt{dBricks} branch are reported to the system domain only, avoiding spurious notifications to downstream teams for revisions that have not yet been released.

Each notification contains general information explaining why the user has been addressed, a link to the corresponding pipeline execution, and a structured table summarizing the detected changes. The table includes the change status (added, modified, or removed) and the previous and current values of each affected element. For every modified or added element, the table includes a direct link to the corresponding entry in \emph{dBricks}, with preconfigured search and filter parameters so that the user can inspect the element in the Single Source of Truth with minimal effort. The link-generation step also serves as a consistency check: if an element is present in the interface data but cannot be resolved to an entry in \emph{dBricks}, the notification is annotated with an explicit warning.

The same change information is reused to populate and update the descriptions of the merge requests that bring revisions from the \texttt{dBricks} branch into \texttt{dev} and from \texttt{dev} into \texttt{master}. Combined with GitLab's native merge-request review workflow, this makes the change information traceable beyond individual email inboxes while keeping the manual effort required of reviewers low. The automation of merge-request generation keeps the system domain's workload low prior to each release.

Figure~\ref{fig:interfacePipeline} illustrates the resulting behavior for a short example case, in which three successive interface revisions are exported from \emph{dBricks}, released to \texttt{dev} in two separate merge requests, and eventually released to \texttt{master}. The example also shows how the notification recipients differ depending on the accumulated change content: the first release to \texttt{dev} carries only a transport-layer code update and notifies the embedded domain only, whereas the second release carries both a function-parameter and a data-type change and therefore notifies both the functional and embedded domains. The subsequent merge into \texttt{master} notifies the same two domains, prompting the downstream repositories to update their submodule references to the final, flight-ready version.

Together, these mechanisms address four of the failure modes identified in Section~\ref{section:failureModes}. \textbf{FM1} is resolved by providing a differential, role-specific overview of each interface change, replacing the manual comparison of regenerated dummy models that was previously required. \textbf{FM2} is addressed in part through the version-controlled XML releases and the proactive notification of downstream teams; the remaining efficiency gain is realized in the software integration repository, where revisions are compared before the time-consuming build steps are triggered (Section~\ref{section:swIntegrationRepo}). \textbf{FM3} is not addressed by the repository structure directly, but the system domain has adopted the convention of defining all function parameters using SI units and handling any required scaling or data-type conversion in the automatically generated transport-layer code. Because both the ICD XML and the Simulink\textsuperscript{\textregistered} dummy models are generated from \emph{dBricks} rather than authored manually, any parameter flowing through the repository inherits the convention by construction. Finally, \textbf{FM5} is addressed by making the system-generated XML the only version that can be committed to the repository. Ad-hoc fixes during bringup must ultimately be made in \emph{dBricks} so that the next automatic export no longer overwrites them silently; local edits to a working copy remain permissible as a debugging aid, but are excluded from the repository by construction and must be reproduced in \emph{dBricks} before they propagate downstream.

\subsection{Function Integration Repository (Functional Domain)}\label{section:functionalRepo}

The function integration repository hosts the Simulink\textsuperscript{\textregistered} integration model of a single integrated function, assembled and maintained through \emph{mrails}. The concept of a dedicated integration repository is not new and has been covered extensively in~\cite{Hochstrasser.Modularmodelbaseddevelopment, Dmitriev.ALeanand}. The contribution of the present work lies in extending the associated pipelines with mechanisms that enforce consistency with the Simulink\textsuperscript{\textregistered} dummy models published by the interface repository, and this subsection therefore focuses on those additions.

Like the interface repository, the function integration repository is organized into three branches. The role of the \texttt{dBricks} mirror branch is played by individual feature branches, forked from \texttt{dev} for each development activity and merged back once the feature has been tested. Interface updates are handled in the same way: a feature branch is opened in response to a notification from the interface repository and is used to perform the corresponding submodule update. Because functional development typically proceeds at a different cadence than interface evolution and several feature branches can be active concurrently, interface updates are not pushed into new feature branches automatically. Instead, the designated integration engineer decides when and how the update is best introduced.

Two automated consistency checks run on every commit to a feature branch and on every merge request into \texttt{dev} or \texttt{master}. The first check verifies \emph{interface-to-model consistency} with a custom MATLAB routine that extracts the inport and outport definitions from the Simulink\textsuperscript{\textregistered} integration model and verifies them against the ICD XML in the referenced interface submodule. This check ensures that port names, directions, and data types are consistent with the interface definitions. The second check ensures \emph{data-dictionary consistency}, and verifies that the contents of the associated Simulink\textsuperscript{\textregistered} Data Dictionary match the referenced definitions.

For merge requests, the pipeline proceeds with code generation using \emph{mrails} after consistency checks have passed, and the merge is blocked until code generation is successful. This guarantees that any model reaching \texttt{dev} or \texttt{master} is known to generate code without errors. On feature-branch commits, by contrast, code generation runs as a non-blocking \emph{dry-run}: failures are reported but do not prevent the commit. The trade-off is deliberate and allows the embedded team to build a new binary from the latest model as soon as it is available. 

The outcomes of both the consistency checks and the dry-run code generation are attached to the descriptions of the automatically generated merge requests so that human review and approval happen against a transparent and traceable record. The same merge-request mechanism is used in the reverse direction when the functional team introduces new parameters or requests changes to the interface definition. In these cases, the merge request is routed upstream to the system domain rather than into \texttt{dev}, providing a standardized channel for cross-domain change proposals.

These additions resolve or mitigate four of the failure modes identified in Section~\ref{section:failureModes}. \textbf{FM1} is addressed by using the notification from the interface repository as the trigger for initiating interface updates in the functional repository. Because the notification carries the structured change summary described above, the functional team starts from a clear differential view rather than having to diff regenerated dummy models manually. The same mechanism also mitigates \textbf{FM2}: because the functional team is notified only of changes that affect the functional interfaces, it can keep working against the latest relevant interface revision while the embedded domain updates its own submodule references at its own cadence, without the two sides drifting silently apart. \textbf{FM4} is addressed by the automated interface and data-dictionary consistency checks, which surface structural mismatches directly in the function integration pipeline rather than during the embedded team's code generation. The dry-run code generation complements this by exposing latent errors in the models. Finally, \textbf{FM6} is addressed by combining the interface-to-model consistency check with the merge-request mechanism. A parameter introduced in the integration model without a counterpart in the referenced ICD XML is flagged by the check, which blocks the merge into \texttt{dev} or \texttt{master}. The functional team therefore cannot silently release such a parameter, and it must either be removed or formally proposed to the system domain, where the addition is consolidated in \emph{dBricks}.

\subsection{Software Integration Repository (Embedded Domain)}\label{section:swIntegrationRepo}

\begin{figure*}[ht]
    \centering
    \includegraphics[width=0.99\textwidth]{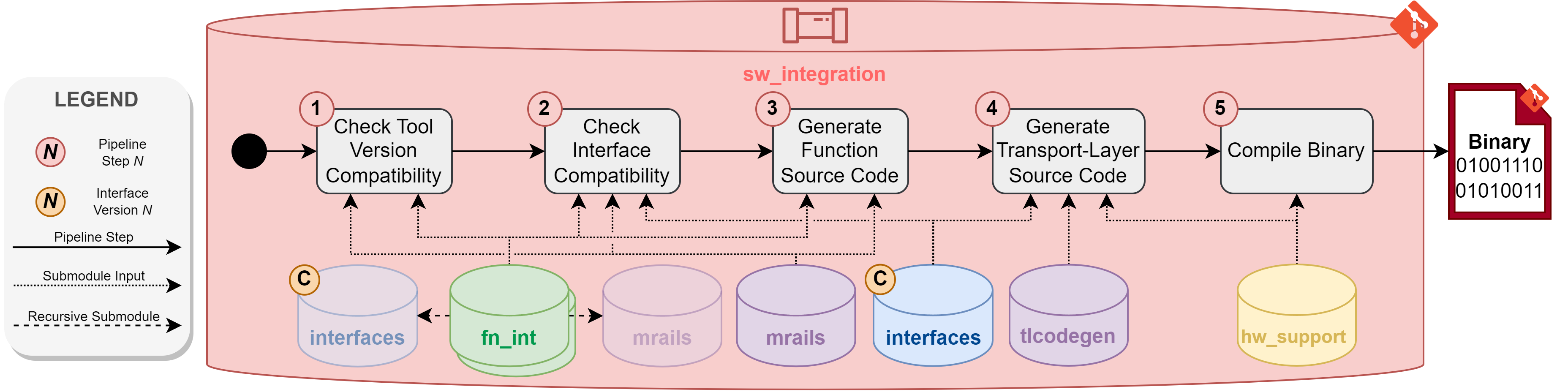}
    \caption{Software Integration Repository Pipeline}\label{fig:swIntegrationPipeline}
\end{figure*}

The software integration repository is the final stop before compilation of the binary. It is typically set up by the embedded team once at the beginning of a project and subsequently used to generate binaries from the latest functional models and interface definitions. To this end, the repository references the interface repository and all function integration repositories. Additionally, a device-specific hardware support repository is also included as a submodule. The low-level software changes comparatively infrequently, and its interaction with the generated code can be largely automated. The embedded domain's manually authored source code is therefore kept in this repository, rather than in a separate repository of its own. The branch structure is identical to the one used in the function integration repository (Section~\ref{section:functionalRepo}).

The associated pipeline is shown in Figure~\ref{fig:swIntegrationPipeline}. Because the pipeline is the sole mechanism by which the testing and flight-ready binary is produced, inconsistencies in any of its inputs could have safety implications, and the pipeline is therefore strictly blocking on failure. It runs on every commit to a feature branch and on every merge request into \texttt{dev} or \texttt{master}, with the same containerized Docker environment used across all pipelines in the proposed approach. The pipeline executes the following five steps:

\begin{enumerate}
    \item \textbf{Tool version compatibility.} For tools that are shared across repositories, such as MATLAB/Simulink\textsuperscript{\textregistered} and \emph{mrails}, the pipeline reads version metadata stored in each referenced submodule and verifies that the versions match the version configured in the software integration repository. Tools used only within this pipeline are versioned through the pipeline's own configuration and therefore do not require cross-repository verification.
    \item \textbf{Interface compatibility.} The same consistency checks introduced in Sections~\ref{section:interfaceRepo} and~\ref{section:functionalRepo} are applied against the \emph{combination} of interface references: the pipeline verifies that the ICD XML of the interface repository referenced by this repository is compatible with the interface revision referenced by each function integration repository. This guarantees that the subsequent code-generation steps consume a consistent view of the interface for all components. The strictness of this check depends on the target branch. Merge requests into \texttt{dev} accept compatible (but not necessarily identical) interface revisions to preserve the flexibility needed during integration and testing. Merge requests into \texttt{master}, by contrast, require an exact version match across all submodules, which produces a clean and reproducible release state.
    \item \textbf{Function source-code generation.} Using \emph{mrails} and the Simulink\textsuperscript{\textregistered} integration model referenced by each function integration repository, functional source code is generated and placed alongside the low-level software.
    \item \textbf{Transport-layer code generation.} Using \emph{tlcodegen}, the referenced ICD XML, and the device-specific templates provided by the hardware support repository, the transport-layer code is generated and placed alongside the low-level software.
    \item \textbf{Binary compilation.} The generated and manually authored sources are compiled into a target-specific binary. The toolchain used is proprietary and GUI-based without a native headless mode; its execution is therefore automated through dedicated Bash and Python wrappers. Device-specific build configurations and auxiliary components required by the compiler are sourced from the hardware support repository.
\end{enumerate}

On successful compilation, the pipeline automatically produces a build report that aggregates the metadata needed to reproduce the build: the versions of the tools used, the Git hashes of all referenced submodules, and checksums of the input artifacts. The compiled binary is exposed as a pipeline artifact and, together with the build report, forms the basis for subsequent integration testing on the target hardware.

These mechanisms address three of the failure modes identified in Section~\ref{section:failureModes}. \textbf{FM2} is addressed in the sense most consequential for development throughput: interface compatibility is verified \emph{before} the time-consuming code-generation and compilation steps are started, so inconsistent revisions across domains no longer surface at the end of a 20--60\,min Simulink\textsuperscript{\textregistered} build phase as described in the original failure mode. This complements the notification mechanism of Section~\ref{section:interfaceRepo} and the interface checks of Section~\ref{section:functionalRepo}, both of which prevent the inconsistency from arising in the first place. \textbf{FM5} is addressed structurally by consuming the interface repository as a referenced submodule rather than maintaining a local copy of the ICD XML. Local ad-hoc edits are not possible within this repository, and any correction made during bringup must be propagated through \emph{dBricks} and the interface repository's release process. Finally, \textbf{FM7} is addressed jointly by the containerized Docker environment, which fixes the execution environment used by every build, and by the tool version compatibility check in Step~1, which verifies that the fixed environment is consistent with the environments against which the upstream artifacts were authored. Together these mechanisms remove the dependence on any individual engineer's locally maintained toolchain.

\subsection{Supporting Repositories}\label{section:supportingRepos}

Alongside the primary repositories, the proposed approach relies on a set of supporting repositories that are frequently reused across projects. These do not introduce new consistency mechanisms of their own, but they are referenced as submodules by the primary repositories and therefore participate in the consistency-checking machinery described in the preceding sections. The supporting repositories fall into two categories.

\textbf{Tools:} Several in-house tools support the described development process, most notably \emph{mrails} and \emph{tlcodegen}. Both are under active development and are referenced as submodules, so that every pipeline invocation runs with matching tool versions to guarantee reproducible code generation results across releases.

\textbf{Hardware Support:} Different hardware platforms require different compilation toolchains and different \emph{tlcodegen} templates. To keep the primary pipelines platform-independent, these hardware-specific artifacts are collected in dedicated hardware support repositories. A hardware support repository may itself reference further submodules such as vendor-provided board support packages or compilation tooling, depending on the complexity of the platform.
\section{Application in Practice}\label{section:application}

The proposed approach is currently being implemented and validated within a manned light sport aircraft project~\cite{Schlautmann.RunTimeAssuranceBased}, with the underlying mechanisms applicable to any airborne system in which the three-domain separation holds. The elements of the proposed approach that promise the greatest benefit have been prioritized for implementation. These include the interface repository, which is fully implemented as described in Section~\ref{section:interfaceRepo}. Additionally, the interface validation stages of the function integration and software integration pipelines have been implemented. The remaining pipeline stages are in progress and the full architecture is not yet in production use.

Initial qualitative feedback from the development team indicates that the implemented mechanisms already reduce inconsistencies arising from interface updates and improve productivity. Members of the functional domain in particular report that the email-based notification system substantially reduces the manual work previously required to identify relevant interface changes. On the system side, exporting the ICD from \emph{dBricks} is now automated through the interface repository's pipeline, which eliminates a manual step that had been a recurring source of delay. A quantitative evaluation of efficiency gains is deferred to future work as broader adoption produces a sufficient sample.
\section{Conclusions and Future Work}\label{section:conclusionFutureWork}

In this paper, a number of failure modes observed during the application of a model-based airborne software development process were identified. Although the underlying processes and corresponding artifacts are well defined, inconsistencies arising between the system, functional, and embedded domains emerge as a bottleneck preventing further productivity gains. These inconsistencies are addressed by a repository-centered approach that defines clear responsibilities and lightweight consistency-checking mechanisms, elevating the repositories from passive storage to central components of the development process through which the interactions between domains are enforced. Choosing lightweight enforcement over heavyweight process gates allows small, resource-constrained teams — such as ours at the Institute of Flight System Dynamics — to benefit from cross-domain consistency without sacrificing agility in the development context.

This paper has presented the proposed approach in detail, including the roles of each repository and the consistency-checking mechanisms of each pipeline. While initial qualitative feedback is promising, the suitability and reliability of the approach must still be established through broader application in real development projects. It is expected that vulnerabilities and weaknesses will surface during broader application and be addressed in subsequent revisions.

Beyond the current implementation, several improvements to the proposed approach are already being considered. First, round-trip time during integration testing could be minimized by automating a full software integration pipeline run as soon as a new commit lands on the corresponding function integration repository. Second, baselining matching interface versions across all repositories for flight testing could be orchestrated through a single automated action. Third, the consistency checks can be extended to cover inter-device communication between independently developed custom devices against a common aircraft-level interface revision. More far-reaching potential lies in a closer integration of architecture models, which would eliminate the manual transfer into \emph{dBricks} and enable derivation of simulation models directly from the architectural description.
\bibliographystyle{IEEEtran}
\bibliography{IEEEabrv,references.bib}
\section{ACKNOWLEDGMENT}

AI tools were used exclusively for language editing (grammar and readability). All technical content and conclusions are the authors' own.

\end{document}